\documentclass[preprint,12pt]{elsarticle}
\usepackage{amssymb,graphics}
\journal{the arXiv.org}

\begin{document}

\begin{frontmatter}

\title{The generating functional for the electromagnetic interaction in the strong gravitational field}

\author{Yuriy Ostapov}
\address{Institute of Mathematics of NAS of Ukraine, 3 Tereschenkivska st.,  Kiev, Ukraine 01601.
E-mail: yugo.ost@gmail.com }

\begin{abstract}
This article is devoted problems of electromagnetic interaction  in curved spacetime. 
Such problems exist, in particular, when we investigate electromagnetic quantum processes near black holes.
The generalization of reduction formalism permits to find formulas for scattering matrices.
For the free Dirac and electromagnetic  fields corresponding generating functionals are calculated.
Next we have found the generating functional for the interaction of these fields.
This result holds the central position in our investigation. 
By means of generating functionals and reduction formulas  we have obtained the scattering amplitudes for elementary electrodynamic processes:
Compton scattering and the annihilation of electron-positron pair (for the tree-level approximation). 
On the base of these results we have formulated the generalized Feynman rules for the electromagnetic interaction in curved spacetime. 
Another electrodynamic processes can be studied by means of these Feynman rules and crossing-symmetry.
The generating functional can be used to study problems of quantum statistics in curved spacetime.

\end{abstract}

\end{frontmatter}

\section{Introduction}

The tendency to unification of quantum field theory with gravity theory obtained real incarnation as yet 87 years ago.
The generalization of the Dirac equation for curved spacetime was considered in the pioneer works Fock V.A.\& Ivanenko D.D. (1929), Schr\"{o}dinger E. (1932).
Investigations of Schr\"{o}dinger E.(1939), DeWitt B.S.(1953), Takahashi Y.\& Umezawa H.(1957), Imamura T.(1960), Parker  L.(1966) et al.  were devoted to particle  creation due to curvature of spacetime.
DeWitt B.S.(1965), Bunch T.S.\& Parker  L.(1979) had found the expansion for the propagators of the scalar and Dirac fields.
The thermal emission for black holes was discovered by Hawking S.W. in 1974 and investigated by Hawking S.W.(1975), DeWitt B.S.(1975), Hartle J.B.(in common with Hawking S.W. in 1976), Christensen S.M.\& Fulling S.A (1977) et al.
\footnote {There is the comprehensive bibliography concerning problems of quantum field theory in curved spacetime (see \cite {Bir,Park}).}.

A lot of articles were devoted to problems of interaction for quantum fields in curved spacetime.
Questions of forming S-matrices, the interaction of scalar fields (including renormalization) as well as particle production due to this interaction are considered in \cite {Bir} ch.9.
The massless theory of scalar field  in asymptotically flat spacetime is studied in \cite {Vol}.
Renormalization in the massless quantum electrodynamics (in de Sitter space) is analysed in \cite {Br,Dr,Sh}.
The first part of \cite {Buch} contains  quantum electrodynamics in curved spacetime  taking into account  particle production in the strong gravity field.
The second part describes quantum processes in conformally flat spacetime.

The given paper is devoted to problems of quantum electrodynamics in curved spacetime: to the interaction of photons with fermions. We shall consider
 Compton scattering and the annihilation of electron-positron pair.
Such problems exist, in particular, when we investigate electromagnetic quantum processes near black holes \cite {Frol}.

First of all we have found the corresponding generating functionals for the free Dirac and electromagnetic  fields as well as for the interaction of these fields.
Using these generating functionals and reduction formalism, we have calculated the scattering matrices for elementary processes:
Compton scattering and the annihilation of electron-positron pair (for the tree-level approximation). 
On the base of these results we have formulated the generalized Feynman rules for electrodynamic processes in curved spacetime. 
The generating functional can be applied for problems of quantum statistics in curved spacetime.

\section{Fermions and photons in curved spacetime}

Preparatory to analysing the interaction of photons with fermions, we consider basic concepts of quantum electrodynamics in curved spacetime \cite {Bir,Park}. 
By $g_{ij} (x)$ denote {\it the metric tensor} of curved spacetime, and $g(x)= det (g_{ij}), dV_x = d^{4}x \sqrt{-g}$. {\it A covariant  derivative} 
is designated by $\nabla_{\mu}$. 

Let $\mu(x)$ be the unit measure concentrated at the point $x=0$, that is,

\begin{equation}
  \int \phi(x)d\mu(x)=\phi(0).                                                           
\end{equation}

Then this Dirac unit measure corresponds to the delta-function $\delta(x)$ determined by the linear functional (see \cite {Sch} ch.2):

\begin{equation}
  \int \phi(x)\delta(x) dV_x=\phi(0).                                                           
\end{equation}

We use as well notations of quantum field theory corresponding \cite {Pes}.

\subsection{The Dirac equation}

In theory of the Dirac field the tetrad formalism  is usually used  to go to a system of locally inertial Cartesian coordinates (see \cite {Wein} ch.12). 
 
Let $X$ be a point in curved spacetime. In the neighborhood of this point we determine  {\it local coordinates} $y^{(a)},a = 0, 1, 2, 3$ \footnote{We use Latin indices $a, b, c$ for local coordinates.}.
Then the metric tensor for local coordinates is $\eta_{(a)(b)}$. Quantities

\begin{equation}
   e^{\mu}_{(a)} (X) = (\frac{{\partial}x^{\mu}} {{\partial}y^{(a)}}  )_{x^{\mu}=X^{\mu}} , \; e^{(a)}_{\mu} (X) = (\frac{{\partial}y^{(a)}} {{\partial}x^{\mu}}  )_{x^{\mu}=X^{\mu}}                                                            
\end{equation}
form a tetrad. The tetrad satisfies the relations:

\begin{equation}
   e^{\mu}_{(a)} (X) e^{(a)}_{\nu} (X) = \delta_{\nu}^{\mu}, \;  e_{\mu}^{(a)} (X) e_{(b)}^{\mu} (X) = {\delta}^{a}_{b} ,
\end{equation}
and

\begin{equation}
   e^{\mu}_{(a)} (X) e^{\nu}_{(b)} (X) g_{{\mu}{\nu}}(X)  = \eta_{(a)(b)} , \; e_{\mu}^{(a)} (X) e_{\nu}^{(b)} (X)\eta_{(a)(b)} = g_{{\mu}{\nu}}(X).                                                        
\end{equation}

As indicated in \cite {Wein} ch.12, {\it the covariant derivative of the spinor field} $\psi(x)$ is

\begin{equation}
 \tilde {\nabla}_{\nu}  {\psi} (x) = [{\partial}_{\nu} + {\Gamma}_{\nu}(x)]{\psi(x)},
\end{equation}
where

\begin{equation}
 {\Gamma}_{\nu} (x)= \frac {1}{8} {\omega}_{(a)(b)\nu}(x)[{\gamma}^{(a)},{\gamma}^{(b)}].
\end{equation}

The coefficients ${\omega}_{(a)(b)\nu}(x)$ are the components for {\it the spin connection}. It can be indicated that

\begin{equation}
  {\omega}_{(a)(b)\nu}(x)= {e_{(a)}^{\mu}}(x){e_{(b)\mu;\nu}}(x).
\end{equation}

The covariant derivative $\tilde {\nabla}_{(a)}$ can be calculated by means of the expression

\begin{equation}
   \tilde {\nabla}_{(a)} = {e_{(a)}^{\mu}}(x) \tilde {\nabla}_{\mu}.
\end{equation}

The action $S$ for the Dirac field is described as

\begin{equation}
   S = \int {\cal L} (x) dV_x ,
\end{equation}
where

\begin{eqnarray}
   {\cal L} (x) &  = &   i {\bar\psi}\gamma^{(a)} e_{(a)}^{\mu}\tilde {\nabla}_{\mu} \psi   - m {\bar\psi}\psi =  \nonumber\\
  &  & =  i {\bar\psi}\tilde{\gamma^{\mu}} (x) \tilde {\nabla}_{\mu} \psi  - m {\bar\psi}\psi.
\end{eqnarray}

In the expression (11) we utilize a set of matrices

\begin{equation}
 \tilde{\gamma^{\mu}} (x) = {e_{(a)}^{\mu}}(x) \gamma^{(a)}.
\end{equation}

Using the variation of the action $S$, we derive {\it the covariant Dirac equation}

\begin{equation}
   i\tilde{\gamma^{\mu}} (x) 
  {\tilde {\nabla}_{\mu}} \psi - m\psi = 0.
\end{equation}

There exists a  set of  solutions $U_{ps}(x), V_{ps}(x)$ of the equation (13) for a fermion with  momentum $p$ and  spin $s$. 
Then we can  use the next expansions for $\psi(x)$ and $\bar{\psi}(x)$:

\begin{equation}
   \psi(x) = \int  d^{3}p \sum_{s} [b(p,s)U_{ps}(x) + d^{\dagger}(p,s)V_{ps}(x)],
\end{equation}

\begin{equation}
  \bar{\psi}(x) = \int  d^{3}p \sum_{s} [b^{\dagger}(p,s){\bar U}_{ps}(x) + d(p,s){\bar V}_{ps}(x)].
\end{equation}

The operators  $b^{\dagger }(p,s), d^{\dagger}(p,s)$ are {\it creation operators} for electrons and positrons respectively, and the operators
$b(p,s), d(p,s)$ represent corresponding {\it annihilation operators}. 

{\it The Green function} $S_{F}(x-x')$ of the Dirac field in curved spacetime is defined as

\begin{equation}
  iS_{F}(x-x') = < 0| T \psi (x) \bar{\psi}(x')|0> 
\end{equation}
and satisfies the following equation

\begin{equation}
  [i \tilde{\gamma^{\mu}} (x) \tilde {\nabla}_{\mu} - m] S_{F}(x-x')= \delta(x-x').
\end{equation}

\subsection{The electromagnetic field}

By $A_{\nu}$ denote the {\it vector potential} of the electromagnetic field.
The action $S$ for the electromagnetic field is described using the Maxwell field strength tensor $F_{\mu \nu}$:

\begin{equation}
   S = \int {\cal L}dV_x,
\end{equation}
where

\begin{equation}
    {\cal L} = - \frac{1}{4} F_{\mu \nu}F^{\mu \nu},
\end{equation}

\begin{equation}
     F_{\mu \nu}= A_{\nu;\mu} - A_{\mu;\nu},\, F^{\mu \nu}= A^{\nu;\mu} - A^{\mu;\nu}.
\end{equation}

Variation of the action $S$ yields the condition\footnote {See \cite {Mis} sec.22.4.}

\begin{equation}
     F^{\mu \nu}{ } { }_{;\nu} = 0.
\end{equation}

Consequently,   it can write

\begin{eqnarray}
     F^{\mu \nu}{ } { }_{;\nu} = {\nabla}_{\nu} {\nabla}^{\mu} A^{\nu} - {\nabla}_{\nu} {\nabla}^{\nu} A^{\mu} = \nonumber\\
= {\nabla}^{\mu} {\nabla}_{\nu} A^{\nu} + R^{\mu}{ }_{\nu}A^{\nu} - {\nabla}_{\nu} {\nabla}^{\nu} A^{\mu} = 0,
\end{eqnarray}
where $R^{\mu}{ }_{\nu}{ }$ is {\it the Ricci tensor} \footnote {Ibid.}. In the Lorentz gauge ${\nabla}_{\nu} A^{\nu}=0$ , and we obtain the equation

\begin{equation}
   {\nabla}_{\nu} {\nabla}^{\nu} A^{\mu} - R^{\mu}{ }_{\nu} A^{\nu}= 0.
\end{equation}

By ${\Re}^{\nu}$ denote {\it the operator De Rham} that satisfies the relation

\begin{equation}
{\Re}^{\nu}A = - {\nabla}_{\mu} {\nabla}^{\mu}A^{\nu} + R^{\nu}{ }_{\mu} A^{\mu}.
\end{equation}

Let $A_{k\lambda}^{\alpha}(x)$ be a  set of solutions of the equation (23) for a photon with  momentum $k$ 
and  polarization $\lambda$.  
 
Then the expansion for $A^{\alpha} (x)$ has the form

\begin{equation}
  A^{\alpha} (x) = \int  d^{3}k \sum_{\lambda} [a_k(\lambda) A^{\alpha}_{k\lambda}(x) + a^{\dagger}_k(\lambda) A^{\alpha\ast}_{k\lambda} (x)].
\end{equation}

We can interpret $a_{k}(\lambda)$ and $a^{\dagger}_{k}(\lambda)$ as annihilation and create operators of photons.

 The Green function $D_{F}^{\mu\nu} (x-x')$ of the electromagnetic field in curved spacetime is determined as

\begin{equation}
 i D^{\mu\nu}_{F}(x-x') = < 0| T A^{\mu} (x)A^{\nu} (x')|0> 
\end{equation}
and satisfies the following equation (in the Lorentz gauge):

\begin{equation}
 - {\nabla}_{\rho} {\nabla}^{\rho} D_{F}^{\mu\nu} (x-x') + R^{\mu}_{\rho} D_{F}^{\rho\nu} (x-x')=\delta(x-x').
\end{equation}

\section{Reduction formalism}

The Lagrangian density for interacting fields takes the form

\begin{eqnarray}
{\cal L} & =  & - \frac{1}{4} F_{\mu \nu}F^{\mu \nu} + i {\bar\psi}\tilde{\gamma^{\mu}} (x) \tilde {\nabla}_{\mu} \psi    - m {\bar\psi}\psi-\nonumber \\
& &  -e\bar{\psi} \tilde {\gamma}^{\mu} (x)\psi A_{\mu}.
\end{eqnarray}

The latter term in (28) describes the interaction of the electromagnetic field with fermions. In- and out-states of fermions are determined as

\begin{equation}
 \lim\limits_{x^0 \to -\infty} \psi(x) = \psi_{in}(x),
\end{equation}

\begin{equation}
 \lim\limits_{x^0 \to \infty} \psi(x) = \psi_{out}(x).
\end{equation}

Using  (14) and (15), we obtain the expansions for  $\psi_{in}(x)$  and  $\bar{\psi}_{in}(x)$:
 
\begin{equation}
   \psi_{in} (x) = \int  d^{3}p \sum_{s} [b_{in}(p,s)U_{ps}(x) + d^{\dagger}_{in}(p,s)V_{ps}(x)],
\end{equation}

\begin{equation}
   \bar {\psi}_{in} (x) = \int  d^{3}p \sum_{s} [b^{\dagger}_{in}(p,s)\bar{U}_{ps}(x) + d_{in}(p,s)\bar{V}_{ps}(x)].
\end{equation}

 By means of the production operators $b^{\dagger}_{in}(p,s)$ and $d^{\dagger}_{in}(p,s)$ we can form an arbitrary in-state. 
Similar expansions can be produced for out-states\footnote{In general, the solutions $U_{ps}(x), V_{ps}(x)$ for out-states will not be equal to the corresponding solutions for in-states.}. 
Replacing in \cite {Bjo} sec.16.9 

\begin{equation}
  {\partial}_ {\mu} \rightarrow \tilde{ {\nabla}_{\mu}},  \,  {\gamma}^{\mu}  \rightarrow \tilde{  {\gamma}^{\mu}},
\end{equation}
we find that

\begin{eqnarray}
   <{\beta} \, out | (ps), {\alpha} \, in>   =  <{\beta} - (ps)out | \alpha \, in>  - \nonumber\\
    - \frac{i}{\sqrt{Z_2}} \int  dV_x < \beta \, out| \bar {\psi}(x) | \alpha \, in> \stackrel {\longleftarrow} {( -i  \tilde {\gamma} \cdot  \tilde {\nabla} -m )} U_{ps}. 
\end{eqnarray}

The first member describes the contribution  only for the elastic scattering. The second member defines  the amplitude of inelastic scattering.
If the in-state contains an antiparticle, then we state for the amplitude of inelastic scattering

\begin{equation}
   \frac{i}{\sqrt{Z_2}} \int  dV_x \, {\overline {V}}_{ps} \stackrel {\longrightarrow} {( i \tilde {\gamma} \cdot  \tilde {\nabla}  -m )} < \beta \, out|  \psi (x) | \alpha \, in>.
\end{equation}

For out-states of particles and antiparticles we have respectively:

\begin{equation}
   - \frac{i}{\sqrt{Z_2}} \int dV_x \, {\overline {U}}_{p's'} \stackrel {\longrightarrow} {( i  \tilde {\gamma} \cdot  \tilde {\nabla} -m )} < \beta \, out|  \psi (x) | \alpha \, in>,
\end{equation}

\begin{equation}
   \frac{i}{\sqrt{Z_2}} \int  dV_x < \beta \, out| \bar {\psi}(x) | \alpha \, in> \stackrel {\longleftarrow} {( -i \tilde {\gamma} \cdot  \tilde {\nabla} -m )} V_{p's'} .
\end{equation}

Consider now the electromagnetic field. The expansion for an in-state takes the form

\begin{equation}
  A^{\alpha}_{in}  (x) = \int  d^{3}k \sum_{\lambda} [a_{in}(k,\lambda) A^{\alpha}_{k\lambda}(x) + a_{in}^{\dagger}(k, \lambda) A^{\alpha\ast}_{k\lambda}(x)]
\end{equation}

By means of the operators $a_{in}^{\dagger}(k,\lambda)$ we can form an arbitrary in-state with n-photons.

The reduction formula for removing a photon from an out-state is \footnote { We have substituted ${\Re}^{\nu} A$ for $\Box A^{\nu}$ in \cite {Bjo} sec.16.10. }

\begin{eqnarray}
   <{\gamma} (k'{\lambda}' )\, out |\phi(x)  |\alpha \, in>   =  <\gamma \, out | \phi(x)| \alpha - (k'{\lambda}' ) \, in>  - \nonumber\\
  - i\frac{1}{\sqrt{Z_3}} \int dV_y <\gamma \, out |T(A_{\mu}(y) \phi(x))|\alpha \, in> \stackrel{\longleftarrow}{\Re}_y A^{\ast\mu}_{k'\lambda'}(y).
\end{eqnarray}

After we shall remove all particles from in- and out-states, we shall arrive a vacuum state.  Consider,  in particularly, the scattering of photons by electrons.
The S-matrix takes the form

\begin{eqnarray}
    <p's';k'{\lambda}' \, out| ps;k\lambda \, in>  =    \delta_{ij}  - \frac{1}{Z_2 Z_3} \int  dV_x \int  dV_{x'}\int  dV_z \int dV_{z'} \times \nonumber\\
    \times A^{\nu}_{k{\lambda}} (x) \stackrel{\longrightarrow}{\Re}_{x} {\bar U}_{p's'}(z')(i\tilde{\gamma}(z') \cdot \tilde {\nabla}_{z'}  -m)\times \nonumber\\
    <0|T(\psi(z') A_{\mu}(x') \bar \psi(z)   A_{\nu}(x))|0> \stackrel{\longleftarrow}{(-i \tilde{\gamma}(z) \cdot \tilde {\nabla}_{z} -m)} U_{ps}(z) \stackrel{\longleftarrow}{{\Re}_{x'} }A^{\ast \mu}_{k'\lambda'}(x')
\end{eqnarray}

Since we are restricted to the tree-level approximation in this article and therefore do not consider the procedure of renormalization,  we shall take $Z_2=Z_3=1$.
The first term in (40) corresponds to the forward scattering.

\section{Generating functionals}

To build Green functions for quantum fields, we shall utilize generating functionals. {\it The functional derivative} is defined as

\begin{equation}
 \frac {\delta F[f(x)]}{\delta f(y)} = \lim\limits_{\varepsilon \to  0} \frac {F[f(x) + \varepsilon\delta(x-y)] - F[f(x)]}{\varepsilon}.
\end{equation}

Functional differentiation in this section is based on the relation

\begin{equation}
 \frac {\delta f(x)}{\delta f(y)} = \delta(x-y)
\end{equation}
as the consequence from (41).

\subsection{Free fields}

Consider the free Dirac field. The generating functional for the Dirac field can be represented as

\begin{eqnarray}
Z^{D}_0 [\eta, \bar{\eta}] & = & \frac{1}{N} \int {\cal D}\bar{\psi} {\cal D}{\psi} \, exp \{ i \int dV_x \times \nonumber\\
& & \times  [\bar{\psi} (x)(i \tilde {\gamma} \cdot \tilde {\nabla} -m ) \psi(x) + \bar{\eta}(x)\psi(x) + \bar{\psi}(x)\eta(x)]\},
\end{eqnarray}
where $\bar{\eta}(x)$ is a source for the field $\psi(x)$, $\eta(x)$ is a source for the field $\bar{\psi}(x)$, N is a number.

Let  S be the operator determined as

\begin{equation}
S^{-1} = i \tilde {\gamma} \cdot \tilde {\nabla} -m .
\end{equation}

Then

\begin{eqnarray}
Z^{D}_0 [\eta, \bar{\eta}]=  \frac{1}{N} \int {\cal D}\bar{\psi} {\cal D}{\psi} \, exp [ i \int dV_x  (\bar{\psi}S^{-1} \psi + \bar{\eta}\psi + \bar{\psi}\eta)].
\end{eqnarray}

Using reasoning in \cite {Ryd} sec. 6.7, we derive \footnote {Taking into account that the normalization condition is $Z^{D}_{0}[0,0]=1$.}

\begin{equation}
Z^{D}_{0} [\eta, \bar{\eta}] = exp[- i\int dV_x dV_y \bar{\eta}(x)S(x-y) \eta(y)]. 
\end{equation}

To find {\it the Dirac propagator}, we carry out functional differentiation:

\begin{eqnarray}
\tau (x,y)  = - {\bigg [\frac {{\delta}^2 Z^{D}_0 [\eta, \bar{\eta}]} {\delta \bar {\eta}(x) \delta {\eta}(y)}\bigg ]}_{ \eta = \bar {\eta} = 0 } =  - \frac{\delta}{\delta \bar {\eta} (x)}\frac{\delta}{\delta {\eta} (y)}\nonumber\\
exp { \bigg \{ - i \int  dV_{x'} dV_{y'}\bar{\eta}(x')S(x'-y') \eta(y') \bigg \} }_{ \eta = \bar {\eta} = 0 } = i S (x-y).
\end{eqnarray}

Since

\begin{equation}
- {\bigg [\frac {{\delta}^2 Z^{D}_0 [\eta, \bar{\eta}]} {\delta \bar{\eta}(x) \delta {\eta}(y)}\bigg ]}_{ \eta = \bar {\eta} = 0 } = <0|T \psi(x)\bar {\psi}(y)|0> = i S_F (x-y),
\end{equation}
then $S_F(x-y)=S(x-y)$, and

\begin{equation}
Z^{D}_{0} [\eta, \bar{\eta}] = exp[- i\int dV_x dV_y \bar{\eta}(x)S_F(x-y) \eta(y)]. 
\end{equation}

Now we can move on to the electromagnetic field. The generating functional for such field can be represented as

\begin{eqnarray}
Z^{E}_0 [J] & = & \frac{1}{N} \int  {\cal D} A \, exp \{ i \int dV_x [ - \frac{1}{4}  F_{\mu \nu}F^{\mu \nu}+ J_{\nu}A^{\nu}]\},
%& & \times  
\end{eqnarray}
where $J$ is a source for the field $A(x)$, N is a number.

At first we shall transform  the integral of the action:

\begin{eqnarray}
S =  -\frac{1}{4} \int dV_x   F_{\mu \nu}F^{\mu \nu} = \nonumber \\
= - \frac{1}{4}\int dV_x    [ {\nabla}_{\mu} A_{\nu} -  {\nabla}_{\nu} A_{\mu} ] [ {\nabla}^{\mu} A^{\nu} -  {\nabla}^{\nu} A^{\mu} ]. 
\end{eqnarray}

After integration by parts \footnote {The covariant form  of the Gauss theorem is described in \cite {Wein}: the formula (4.7.8). Besides we use the relation (16.6) from \cite {Mis}.}:

\begin{eqnarray}
S =  \frac{1}{2}\int dV_x   [ A_{\nu}({\nabla}_{\mu} {\nabla}^{\mu}A^{\nu} -  {\nabla}_{\mu}{\nabla}^{\nu} A^{\mu}) ] = \nonumber \\
=  \frac{1}{2}\int dV_x  A_{\nu}({\nabla}_{\mu} {\nabla}^{\mu}A^{\nu} -  {\nabla}^{\nu}{\nabla}_{\mu} A^{\mu} - R_{\mu}^{\nu} A^{\mu}).
\end{eqnarray}

Using the Lorentz gauge, we state the next relation

\begin{eqnarray}
S =  -\frac{1}{4} \int dV_x    F_{\mu \nu}F^{\mu \nu} = \nonumber \\
=  \frac{1}{2}\int dV_x   A_{\nu}({\nabla}_{\mu} {\nabla}^{\mu}A^{\nu} - R_{\mu}^{\nu} A^{\mu}).
\end{eqnarray}

After  substitution $S$ in (50) the  generation functional for the electromagnetic field gets the form

\begin{eqnarray}
Z^{E}_0 [J]  = \frac{1}{N} \int  {\cal D} A \, exp \{ i \int dV_x [  \frac{1}{2} A_{\nu}({\nabla}_{\mu} {\nabla}^{\mu}A^{\nu} - R_{\mu}^{\nu} A^{\mu}) + J_{\nu}A^{\nu}]\}.
\end{eqnarray}

Taking into account (24), we obtain

\begin{equation}
\frac{1}{2}A_{\nu} ({\nabla}_{\mu} {\nabla}^{\mu}A^{\nu} - R_{\mu}^{\nu} A^{\mu}) + J_{\nu}A^{\nu} = -\frac{1}{2}A^{T}{\Re}A + J^{T} A.
\end{equation}

Let  D be the operator determined as

\begin{equation}
D = - {\Re}^{-1}.
\end{equation}

We can transform (54) using the functional extension for the relation (6.26) from \cite {Ryd} sec. 6.2 \footnote {We use as well the normalization condition $Z^{E}_0 [0] = 1$.}:

\begin{equation}
Z^{E}_0 [J] = exp \{- \frac{i}{2} \int J^{T}(x) D(x-y) J(y)dV_x dV_y \}.
\end{equation}

 Functional differentiation yields

\begin{eqnarray}
\tau (x,y)  = - {\bigg [\frac {{\delta}^2 Z^{E}_0 [J]} {\delta J(x) \delta J(y)}\bigg ]}_{ J  = 0 } =  \nonumber\\
=  - \frac{\delta}{\delta J (x)}\frac{\delta}{\delta J (y)} exp { \bigg \{ - \frac{i}{2} \int dV_{x_1} dV_{x_2}  J^{T}(x_1)D(x_1-x_2) J(x_2) \bigg \} }_{ J  = 0 } = \nonumber\\
= - \frac{\delta}{\delta J (x)} \bigg  [- \frac {i}{2} \int dV_{x_2} D(y-x_2)J(x_2) - \frac {i}{2} \int dV_{x_1}J^{T}(x_1) D(x_1-y) \bigg ]\times \nonumber\\
\times exp { \bigg \{ - \frac{i}{2} \int dV_{x_1} dV_{x_2}  J^{T}(x_1)D(x_1-x_2) J(x_2) \bigg \} }_{ J  = 0 } = iD(x-y)
\end{eqnarray}

Since

\begin{equation}
 - {\bigg [\frac {{\delta}^2 Z^{E}_0 [J]} {\delta J(x) \delta J(y)}\bigg ]}_{ J  = 0 } = <0|T A(x) A(y)|0> = i D_F (x-y),
\end{equation}
then $D_F(x-y)=D(x-y)$, and

\begin{equation}
Z^{E}_0 [J] = exp \{- \frac{i}{2} \int J^{T}(x) D_F(x-y) J(y)dV_x dV_y \}.
\end{equation}

\subsection{Interacting fields}

The generating functional for  the interaction of photons with fermions takes the form

\begin{eqnarray}
Z [\eta, \bar{\eta}, J] & = & \frac{1}{N} \int {\cal D}\bar{\psi} {\cal D}{\psi} {\cal D} A\, exp \{ iS_0 + i\int dV_x [J_{\mu}A^{\mu}  + \nonumber\\
& &  + \bar{\eta}(x)\psi(x) + \bar{\psi}(x)\eta(x)] + i\int {\cal L}_{int} d V_x \},
\end{eqnarray}
where

\begin{eqnarray}
S_0  =   \int \{  i \bar{\psi}\tilde{\gamma^{\mu}} (x) \tilde {\nabla}_{\mu} \psi   - m {\bar\psi}\psi  - \frac{1}{4}  F_{\mu \nu}F^{\mu \nu} \} dV_x,
\end{eqnarray}

\begin{equation}
{\cal L}_{int} = -e\bar{\psi} \tilde {\gamma}_{\mu} \psi A^{\mu}.
\end{equation}

The numerator for the generating functional of the free fields $Z_0[\eta , \bar {\eta}, J]$ is determined by the expression 

\begin{eqnarray}
Z_0 [\eta, \bar{\eta}, J] & = & \int {\cal D}\bar{\psi} {\cal D}{\psi} {\cal D} A\, exp \{ iS_0 + i\int dV_x [J_{\mu}A^{\mu}  + \nonumber\\
& &  + \bar{\eta}(x)\psi(x) + \bar{\psi}(x)\eta(x)] \}.
\end{eqnarray}

Using the first three members in the expansion

\begin{equation}
exp (i\int {\cal L}_{int} d V_x) = 1 - ie \int \bar{\psi} \tilde {\gamma}_{\mu} \psi A^{\mu} dV_x + \frac{1}{2!} {(ie \int \bar{\psi} \tilde {\gamma}_{\mu} \psi A^{\mu} dV_x)}^2 +...
\end{equation}
and  substituting in (61), we obtain \footnote{Thereafter we can omit indices for $J, \tilde {\gamma}$, and $D_F$.}

\begin{eqnarray}
Z [\eta, \bar{\eta}, J]  =  \frac{1}{N}  \int {\cal D}\bar{\psi} {\cal D}{\psi} {\cal D} A [1 - (ie \int \bar{\psi} \tilde {\gamma}_{\mu} \psi A^{\mu} dV_{z})+ \frac {1}{2!}{(ie \int \bar{\psi} \tilde {\gamma}_{\mu} \psi A^{\mu} dV_{z})}^2 ] \times \nonumber\\
 \times  exp \{ iS_0 + i\int dV_x [J_{\mu}A^{\mu}   + \bar{\eta}(x)\psi(x) + \bar{\psi}(x)\eta(x)] \} = \nonumber\\
 = \frac{1}{N} \bigg [1 - \bigg (ie \int (\frac {1}{i} \frac {\delta} {\delta \eta}) \tilde {\gamma}(z) (\frac {1}{i} \frac {\delta} {\delta \bar {\eta}}) (\frac {1}{i} \frac {\delta} {\delta J}) dV_{z} \bigg  ) + \nonumber\\
+ \frac {1} {2!} {\bigg  (ie \int (\frac {1}{i} \frac {\delta} {\delta \eta}) \tilde {\gamma}(z) (\frac {1}{i} \frac {\delta} {\delta \bar {\eta}}) (\frac {1}{i} \frac {\delta} {\delta J}) dV_{z}\bigg  )}^2 \bigg  ] Z_0[\eta , \bar {\eta}, J]
\end{eqnarray}

 Let $z_1$ and $z_2$ be the points of collision and decay. Then

\begin{eqnarray}
Z [\eta, \bar{\eta} , J]  =  \frac{1}{N} \bigg [1 - \bigg (ie \int (\frac {1}{i} \frac {\delta} {\delta \eta(z)}) \tilde {\gamma}(z) (\frac {1}{i} \frac {\delta} {\delta \bar {\eta}(z)}) (\frac {1}{i}  \frac {\delta} {\delta J(z)}) dV_{z}\bigg )+ \nonumber\\
 + \frac {1} {2!}\bigg (ie \int (\frac {1}{i} \frac {\delta} {\delta \eta(z_2)}) \tilde {\gamma}(z_2) (\frac {1}{i} \frac {\delta} {\delta \bar {\eta}(z_2)}) (\frac {1}{i}  \frac {\delta} {\delta J(z_2)}) dV_{z_2}\bigg ) \times \nonumber\\
\times \bigg (ie \int (\frac {1}{i} \frac {\delta} {\delta \eta(z_1)}) \tilde {\gamma}(z_1) (\frac {1}{i} \frac {\delta} {\delta \bar {\eta}(z_1)}) (\frac {1}{i}  \frac {\delta} {\delta J(z_1)}) dV_{z_1}\bigg ) \bigg ] Z_0[\eta , \bar {\eta}, J]], 
\end{eqnarray}
where $N$ is equal to the numerator of the relation (67) for $J = \eta = \bar {\eta} = 0$.

To apply this expression, it should be borne in mind that according  to (49) and (60):

\begin{eqnarray}
Z_{0} [\eta, \bar{\eta}, J] = exp\{- i\int dV_x dV_y [\bar{\eta}(x)S_F(x-y) \eta(y) + \frac{1}{2} J(x)D_F(x-y) J(y) ]\}.
\end{eqnarray}

After functional differentiation we have \footnote {We shall write further $exp\{...\}$ instead of $exp\{- i\int dV_x dV_y [\bar{\eta}(x)S_F(x-y) \eta(y) +  \frac{1}{2} J(x)D_F(x-y) J(y) ]\}$ .}

\begin{eqnarray}
(\frac {1}{i}\frac {\delta} {\delta J(z)})Z_0[\eta , \bar {\eta}, J]= -\int D_F(z-x)J(x) dV_x  exp\{...\}   \nonumber \\
(\frac {1}{i} \frac {\delta} {\delta \bar {\eta}(z)})(\frac {1}{i}\frac {\delta} {\delta J(z)})Z_0[\eta , \bar {\eta}, J]= \int D_F(z-x)J(x) dV_x \int S_F(z-y){\eta}(y) dV_y exp\{...\}   \nonumber \\
(\frac {1}{i} \frac {\delta} {\delta {\eta}(z)}) \tilde {\gamma}(z) (\frac {1}{i} \frac {\delta} {\delta \bar {\eta}(z)})(\frac {1}{i} \frac {\delta} {\delta J(z)})Z_0[\eta , \bar {\eta}, J]= \tilde {\gamma}(z)[\frac{1}{i} S_F(0)\int D_F(z-x)J(x)dV_x -\nonumber \\
-  \int D_F(z-x)J(x)dV_x \int S_F (z-y) \eta(y)dV_y \int \bar {\eta}(x) S_F(x-z) dV_x] exp\{...\} 
\end{eqnarray}

Substituting (69) in (67), we state for the members of the first order:

\begin{eqnarray}
Z^{(1)} [\eta, \bar{\eta} , J]  =  ie \int dV_z\ \tilde {\gamma}(z)\{-i S_F(0)\int D_F(z-x)J(x)dV_x  - \nonumber \\
-  \int D_F(z-x)J(x)dV_x \int S_F (z-y) \eta(y)dV_y \int \bar {\eta}(x) S_F(x-z) dV_x\} exp\{...\} 
\end{eqnarray}

We shall establish further in sec.5.1 that the members of the first order do not yield the contribution to the scattering matrix. 
Therefore, we shall consider as well the members of the second order according to (67). Fulfilling the first differentiation, we derive

\begin{eqnarray}
(\frac {1}{i}\frac {\delta} {\delta J(z_2)}) Z^{(1)} [\eta (z_1), \bar{\eta}(z_1) , J(z_1)] = ei\int dV_{z_1} \tilde {\gamma}(z_1) \times \nonumber \\
\times [-  S_F(0) D_F(z_1-z_2) + i \tilde {\gamma}_{\mu} S_F(0) \int D_F(z_1-x)J(x)dV_x \int D_F(z_2-x)J(x)dV_x + \nonumber \\
+ i D_F(z_1-z_2) \int S_F(z_1-y)\eta(y)dV_y \int \bar {\eta} (x) S_F(x-z_1)dV_x + \nonumber \\
+  \int D_F(z_1-x)J(x)dV_x \int S_F(z_1-y)\eta(y)dV_y \int \bar {\eta} (x) S_F(x-z_1)dV_x \times \nonumber \\
\times \int D_F(z_2-x)J(x)dV_x] exp\{...\}
\end{eqnarray}

Using the second differentiation, we have

\begin{eqnarray}
(\frac {1}{i} \frac {\delta} {\delta \bar {\eta}(z_2)})(\frac {1}{i}\frac {\delta} {\delta J(z_2)}) Z^{(1)} [\eta (z_1), \bar{\eta}(z_1) , J(z_1)] = ei\int dV_{z_1} \tilde {\gamma}(z_1)\times  \nonumber \\
\times [- i S_F(0) \int D_F(z_1-x)J(x)dV_x \int D_F(z_2-x)J(x)dV_x \int S_F(z_2-y)\eta(y)dV_y + \nonumber \\
+ S_F(0)D_F(z_1-z_2)\int S_F(z_2-y)\eta(y)dV_y   + \nonumber \\
+ D_F(z_1-z_2)S_F(z_2-z_1)\int S_F(z_1-y)\eta(y)dV_y -\nonumber \\
- i D_F(z_1-z_2) \int S_F(z_1-y)\eta(y)dV_y \int \bar {\eta} (x) S_F(x-z_1)dV_x \int S_F(z_2-y)\eta(y)dV_y - \nonumber \\
- i S_F(z_2-z_1)\int D_F(z_1-x)J(x)dV_x \int S_F(z_1-y)\eta(y)dV_y \int D_F(z_2-x)J(x)dV_x -\nonumber \\
- \int D_F(z_1-x)J(x)dV_x \int S_F(z_1-y)\eta(y)dV_y \int \bar {\eta} (x) S_F(x-z_1)dV_x  \times \nonumber \\
\times \int D_F(z_2-x)J(x)dV_x \int S_F(z_2-y)\eta(y)dV_y] exp\{...\}
\end{eqnarray}

After the third differentiation we obtain

\begin{eqnarray}
(\frac {1}{i} \frac {\delta} {\delta {\eta}(z_2)}) \tilde {\gamma} (z_2) (\frac {1}{i} \frac {\delta} {\delta \bar {\eta}(z_2)})(\frac {1}{i} \frac {\delta} {\delta J(z_2)}) Z^{(1)} [\eta (z_1), \bar{\eta}(z_1) , J(z_1)] =  \nonumber \\
= ei\int dV_{z_1}\tilde {\gamma} (z_2) \tilde {\gamma}(z_1) [ -i S_F(0)D_F(z_1-z_2) S_F(0)-\nonumber \\
- S_F(0)D_F(z_1-z_2)\int S_F(z_2-y)\eta(y)dV_y\int \bar {\eta} (x) S_F(x-z_2)dV_x +\nonumber \\
%2
-  S_F(0) S_F(0)\int D_F(z_1-x)J(x)dV_x \int D_F(z_2-x)J(x)dV_x    + \nonumber \\
+ i S_F(0) \int D_F(z_1-x)J(x)dV_x \int D_F(z_2-x)J(x)dV_x \int S_F(z_2-y)\eta(y)dV_y \times \nonumber \\
 \times \int \bar {\eta} (x) S_F(x-z_2)dV_x - \nonumber \\
%3
  -i  D_F(z_1-z_2)S_F(z_2-z_1) S_F(z_1-z_2) -\nonumber \\
- D_F(z_1-z_2)S_F(z_2-z_1)\int S_F(z_1-y)\eta(y)dV_y\int \bar {\eta} (x) S_F(x-z_2)dV_x -\nonumber \\
%4
- D_F(z_1-z_2)  S_F(z_1-z_2) \int \bar {\eta} (x) S_F(x-z_1)dV_x \int S_F(z_2-y)\eta(y)dV_y - \nonumber \\
- D_F(z_1-z_2) S_F(0)\int S_F(z_1-y)\eta(y)dV_y \int \bar {\eta} (x) S_F(x-z_1)dV_x   + \nonumber \\
+ i D_F(z_1-z_2) \int S_F(z_1-y)\eta(y)dV_y \int \bar {\eta} (x) S_F(x-z_1)dV_x \times \nonumber \\
\times \int S_F(z_2-y)\eta(y)dV_y \int \bar {\eta} (x) S_F(x-z_2)dV_x - \nonumber \\
%5
- S_F(z_2-z_1)S_F(z_1-z_2) \int D_F(z_1-x)J(x)dV_x  \int D_F(z_2-x)J(x)dV_x +\nonumber \\
+ i S_F(z_2-z_1)\int D_F(z_1-x)J(x)dV_x \int S_F(z_1-y)\eta(y)dV_y \times\nonumber \\
\times \int D_F(z_2-x)J(x)dV_x \int \bar {\eta} (x) S_F(x-z_2)dV_x +\nonumber \\
%6
+ i   S_F(z_1-z_2) \int D_F(z_1-x)J(x)dV_x \int \bar {\eta} (x) S_F(x-z_1)dV_x  \times \nonumber \\
\times \int D_F(z_2-x)J(x)dV_x \int S_F(z_2-y)\eta(y)dV_y +\nonumber \\
+i S_F(0)\int D_F(z_1-x)J(x)dV_x \int S_F(z_1-y)\eta(y)dV_y \int \bar {\eta} (x) S_F(x-z_1)dV_x  \times \nonumber \\
\times \int D_F(z_2-x)J(x)dV_x   +\nonumber \\
+ \int D_F(z_1-x)J(x)dV_x \int S_F(z_1-y)\eta(y)dV_y \int \bar {\eta} (x) S_F(x-z_1)dV_x  \times \nonumber \\
\times \int D_F(z_2-x)J(x)dV_x \int S_F(z_2-y)\eta(y)dV_y] \int \bar {\eta} (x) S_F(x-z_2)dV_x ] exp\{...\}
\end{eqnarray}

 Since according to (73) $N\neq 1$, we present $N$ as $N=1 - e^2 B$, where $B$ consists of the corresponding members of the numerator in the expression (67) subtracted from $Z_0$. 
These members have not  integrals containing functions $J,\eta, \bar {\eta }$ before $exp\{...\}$ and therefore are not equal to 0 for $J = \eta = \bar {\eta} = 0$.

Let $A_0$ and $A_1$ be the members of the numerator in (67) containing integrals before $exp\{...\}$. These members vanish for  $J = \eta = \bar {\eta} = 0$.
Then in the second approximation

\begin{eqnarray}
Z [\eta, \bar{\eta} , J]  =  \frac {[1 - ieA_0 - e^2(A_1 + B)] Z_0[\eta , \bar {\eta}, J] }{1 - e^2B}  \approx  \nonumber \\
\approx [1 - ieA_0 - e^2(A_1 + B)](1+e^2B) Z_0[\eta , \bar {\eta}, J] \approx [1 - ieA_0 - e^2A_1 )] Z_0[\eta , \bar {\eta}, J]
\end{eqnarray}

Consequently, the term $B$ vanishes from $Z [\eta, \bar{\eta} , J]$. 

Using (67), (69), and (73) in common with (74), we can in the next section to build the corresponding Green functions for elementary processes.

\section{Elementary processes of quantum electrodynamics}

\subsection{The Green function for the interaction of fermions with photons}

To find the matrix scattering by means of expression (40), we need the Green function $ \tau (x_1, x_2,x_3,x_4)$. Consider the function

\begin{equation}
\tau (x_1, x_2,x_3,x_4) = {\bigg [\frac {\delta} {\delta \bar {\eta}(x_4)} \frac {\delta} {\delta J(x_3)}  \frac {\delta} {\delta  {\eta}(x_2)}\frac {\delta} {\delta J(x_1)} Z [\eta, \bar{\eta}, J] \bigg ]}_ {\eta= \bar {\eta} = J = 0}
\end{equation}
that corresponds to  Compton scattering:

\begin{equation}
e^{-}\gamma \rightarrow e^{-}\gamma 
\end{equation}
where $J (x_1)$ and $J (x_3)$ are sources for the input  and output photons respectively, and  $\eta(x_2)$ and $\bar {\eta}(x_4)$  are sources for the input  and output electrons.  

Consider operations of differentiation  according to (75).
 The result depends on the presence of integrals containing functions $J,\eta, \bar {\eta }$ before $exp\{...\}$ in the members of sums (70) and (73) (after differentiation) 
 \footnote { Integrals containing functions $\eta, \bar {\eta }$  may be only paired: the integral with $\eta$ and the integral with $\bar {\eta}$ .}:

\begin {enumerate}
\item If  integrals with $J$ are absent, then after two  differentiations   of $exp\{...\}$ wrt $J$ the propagator of the free electromagnetic field appears (see sec.4.1)\footnote {Hereafter we shall use the abbreviation "wrt" instead of "with respect to".}. 

\item If there is one integral with $J$, then after  the first differentiation  wrt $J$ it disappears, and after  the second differentiation   of $exp\{...\}$ wrt $J$ the integral with $J$ appears.

\item If there are two integrals with $J$, then these integrals  disappear after two  differentiations wrt $J$.

\item If there are more two integrals with $J$, then all integrals can not disappear.

\item If  integrals containing functions $\eta, \bar {\eta }$ are absent, then after the first and second  differentiation  of $exp\{...\}$ wrt $\eta, \bar {\eta }$ the propagator of the free Dirac field appears (see sec.4.1).

\item If there is one pair of integrals containing functions $\eta, \bar {\eta }$, then these integrals  disappear after the first and second  differentiation wrt $\eta, \bar {\eta }$ .

\item If there are  two pairs of integrals, then after the first and second  differentiation wrt $\eta, \bar {\eta }$ only one pair  disappears.
\end {enumerate}

According to the items 1-7 (taking account of (67), (69), and (73) in common with (74)), we can do such conclusions:

\begin {itemize}
\item the member $Z_0[\eta , \bar {\eta}, J]$ leads to the propagators of free fields;
\item the members from (70) do not be included to the Green function (75);
\item the members from (73) do not be taken into account if they contain $S_F(0)$ and therefore form  loop diagrams;
\item the members including $D_F(z_1 - z_2)S_F(z_1 - z_2)$ do not be taken into account since they describe in common a virtual photon and fermion;
\item the member

\begin{eqnarray}
e^2 \frac {i}{2} \int dV_{z_1} dV_{z_2}\tilde {\gamma}(z_1)\tilde {\gamma}(z_2)S_F(z_2 -z_1) \int D_F(z_1-x)J(x) dV_x \times  \nonumber \\
\times \int S_F(z_1-x) \eta(x) dV_x  \int D_F(z_2-x)J(x) dV_x \int S_F(x-z_2) \bar {\eta}(x) dV_x exp \{...\}
\end{eqnarray}
yields two contributions to the Green function (75):

\begin{eqnarray}
\tau (x_1, x_2,x_3,x_4) = i{e^2} \int dV_{z_1} dV_{z_2} \tilde {\gamma}(z_1) \tilde {\gamma} (z_2) S_F(z_2 -z_1) \times \nonumber \\
\times [D_F(z_1 - x_1)D_F(z_2 - x_3) + D_F(z_1 - x_3)D_F(z_2 - x_1) ]S_F(z_1 - x_2)S_F(x_4 - z_2)  
\end{eqnarray}

\end {itemize}

The factor $\frac{1}{2}$ vanishes since there is another member in (73) that is equal to (77) after the substitution $z_1\leftrightarrow z_2$.

Thus, we receive the result corresponding to  classical quantum electrodynamics. However, instead of usual Green functions of QED it should be used Green functions for curved spacetime.

Now we can move on to calculating the scattering matrix  for elementary quantum electrodynamic processes.

\subsection{The interaction of fermions with photons}

To find the scattering matrix for the reaction (76), we substitute the relation (78) in the expression (40) taking into account that $x_1=x, x_2=z, x_3=x', x_4=z'$ . 
 At first consider the first member  in (78):

\begin{eqnarray}
      -  \int  dV_x   dV_{x'}  dV_z  dV_{z'} A_{k\lambda}^{\mu} (x) \stackrel{\longrightarrow}{{\Re}_x} {\bar U}_{p's'}(z')(i \tilde {\gamma}(z') \cdot  {\nabla}_{z'} -m)\times \nonumber\\
\times  {ie^2} \int dV_{z_1} dV_{z_2} \tilde {\gamma}_{\mu} (z_1)\tilde {\gamma}_{\nu} (z_2) S_F(z_2 -z_1)D_F(z_1 - x)D_F(z_2-x')S_F(z_1 - z)S_F(z'-z_2) \times \nonumber\\
\times \stackrel{\longleftarrow}{(-i\tilde {\gamma}(z) \cdot  {\nabla}_{z} -m)} U_{ps}(z) \stackrel{\longleftarrow}{{\Re}_{x'}} A_{k'\lambda'}^{\nu\ast} (x')
\end{eqnarray}

We have the relations:

\begin{equation}
 D_F(z_2 - x')\stackrel{\longleftarrow}{{\Re}_{x'}}= \delta (z_2 -x'),
\end{equation}

\begin{equation}
  {\Re}_x  D_F(z_1-x) = \delta (z_1 - x),
\end{equation}

\begin{equation}
(i  \tilde {\gamma}(z')\cdot \tilde {\nabla}_{z'} -m)  S_F(z' - z_2)=\delta (z' - z_2),
\end{equation}

\begin{equation}
S_F(z_1-z) \stackrel{\longleftarrow}{(-i \tilde {\gamma}(z) \cdot  {\nabla}_{z}  -m)}= \delta (z_1 - z).
\end{equation}

Taking into account (80) - (83),  the first member has the form

\begin{eqnarray}
     - i{e^2} \int  dV_{z_1} dV_{z_2} \bar {U}_{p's'} (z_2) \tilde {\gamma}_{\nu} (z_2) A^{\ast\nu}_{k'\lambda'}(z_2) S_F(z_2 -z_1)\tilde {\gamma}_{\mu} (z_1) A_{k\lambda}^{\mu} (z_1)  U_{ps} (z_1).
\end{eqnarray}

For the second member in (78) $z_2 =x, z_1 =x'$. Therefore, this member is equal to

\begin{eqnarray}
     - i{e^2} \int  dV_{z_1} dV_{z_2} \bar {U}_{p's'} (z_2) \tilde {\gamma}_{\nu} (z_2) A^{\nu}_{k\lambda}(z_2) S_F(z_2 -z_1)\tilde {\gamma}_{\mu}(z_1) A_{k'\lambda'}^{\ast\mu} (z_1)  U_{ps} (z_1).
\end{eqnarray}

Thus,

\begin{eqnarray}
  S_{fi} =   - i{e^2} \int  dV_{z_1} dV_{z_2}[ \bar {U}_{p's'} (z_2) \tilde {\gamma}_{\nu}(z_2)A^{\ast\nu}_{k'\lambda'}(z_2) S_F(z_2 -z_1)\tilde {\gamma}_{\mu}(z_1) A_{k\lambda}^{\mu} (z_1)  U_{ps} (z_1) + \nonumber\\
+ \bar {U}_{p's'} (z_2) \tilde {\gamma}_{\nu}(z_2) A^{\nu}_{k\lambda}(z_2) S_F(z_2 -z_1)\tilde {\gamma}_{\mu}(z_1) A_{k'\lambda'}^{\ast\mu} (z_1)  U_{ps} (z_1)]
\end{eqnarray}

For  Minkowski space we can present  $U_{ps}, \bar {U}_{p's'},A_{k\lambda}^{\mu}, A^{\ast\nu}_{k'\lambda'}$ in terms of plane waves.
Substituting in (86) 

\begin{equation}
S_F(z_2 -z_1) = \frac {1}{{(2\pi)}^4}\int dq \, exp (-iq(z_2 -z_1)) \tilde S_F(q),
\end{equation}
where $\tilde S_F(q)$ is the Fourier transform for $S_F(z)$, we shall find that all signs of integrals in (86) vanish, and we shall obtain 
the usual expression for Compton scattering  (see \cite {Pes} sec.5.5).

\subsection{The annihilation of electron-positron pair}

The annihilation of electron-positron pair corresponds to the reaction

\begin{equation}
e^{-}e^{+} \rightarrow \mu^{-}\mu^{+}.
\end{equation}

Using reduction formalism, we shall establish that the matrix scattering gets the form

\begin{eqnarray}
    <p's';q'r'  \, out|ps;qr \,in>  =   \delta_{ij} -  \int  dV_x \int  dV_{x'} \int  dV_z \int dV_{z'} \times \nonumber\\
    \times{\bar U}_{p's'}(x')(i \tilde{\gamma}(x') \cdot \tilde {\nabla}_{x'}   -m_{\mu}){\bar V}_{qr}(z)(i \tilde{\gamma}(z) \cdot \tilde {\nabla}_{z}   -m_{e})\times \nonumber\\
    \times <0|T(\bar {\psi}_{{\mu}^{+} }(z')  {\psi}_{{\mu}^{-} }(x')  {\psi}_{e^{+}}(z) \bar {\psi}_{e^{-}}(x))|0>\times \nonumber\\
   \times  \stackrel{\longleftarrow}{(-i \tilde{\gamma}(z')  \cdot \tilde {\nabla}_{z'}   -m_{\mu})} V_{q'r'}(z') \stackrel{\longleftarrow}{(-i \tilde{\gamma}(x)\cdot  \tilde {\nabla}_{x}  -m_{e})} U_{ps}(x).
\end{eqnarray}

To find the matrix scattering by means of the expression (89), we need the Green function $ \tau (x_1, x_2, x_3, x_4)$. Consider the function

\begin{equation}
\tau (x_1, x_2,x_3,x_4) = {\bigg [\frac {\delta} {\delta \bar {\eta}(x_4)} \frac {\delta}  {\delta  {\eta}(x_3)} \frac {\delta} {\delta \eta(x_2)}\frac {\delta} {\delta \bar {\eta}(x_1)} Z [\eta, \bar{\eta}, J] \bigg ]}_ {\eta= \bar {\eta} = J = 0},
\end{equation}
where $\bar {\eta} (x_1)$ and $\eta (x_2)$ are sources for an electron and positron respectively, and  $\eta(x_3)$ and $\bar {\eta}(x_4)$  are sources for muons.

The result in (90) depends on the presence of integrals containing functions $J, \eta, \bar {\eta }$ before $exp\{...\}$ in the members of sums (70) and (73) (after differentiation)
\footnote{ We recall that integrals containing $\eta, \bar {\eta }$ may be only paired. }:
\begin {enumerate}
\item If  integrals containing $\eta, \bar {\eta }$ are absent, then after the first and second  differentiation  of $exp\{...\}$ wrt $\eta, \bar {\eta }$ the propagator of the free Dirac field appears (see sec.4.1). 
After the third  and fourth  differentiation  of $exp\{...\}$ another propagator of the free Dirac field appears.

\item If there is one pair of integrals containing $\eta, \bar {\eta }$, then these integrals  disappear after the first and second  differentiation wrt $\eta, \bar {\eta }$.
 After the third  and fourth  differentiation  of $exp\{...\}$ the propagator of the free Dirac field appears.

\item If there are  two pairs of integrals, then all integrals  disappear  after all differentiations wrt $\eta, \bar {\eta }$.
\end {enumerate}

According to the items 1-3 (using (67), (69), and (73) in common with (74)) we can do such conclusions: only the member

\begin{eqnarray}
e^2 \frac {i}{2} \int dV_{z_1}dV_{z_2}\tilde {\gamma}(z_2) \tilde {\gamma} (z_1)D_F(z_1-z_2) \int S_F(z_1-y)\eta(y)dV_y \int \bar {\eta} (x) S_F(x-z_1)dV_x \times \nonumber \\
\times \int S_F(z_2-y)\eta(y)dV_y \int \bar {\eta} (x) S_F(x-z_2)dV_x
\end{eqnarray}
gives the contribution to the Green function:

\begin{eqnarray}
\tau (x_1, x_2,x_3,x_4) = \frac{ie^2}{2}\int dV_{z_1}dV_{z_2}\tilde {\gamma}(z_2)\tilde {\gamma}(z_1) D_F(z_2 - z_1) [S_F( z_1 - x_2 )S_F( x_1 -z_1 )\times \nonumber \\
\times S_F(z_2 - x_3)  S_F(x_4- z_2) +  S_F( z_2 - x_2 )S_F( x_1 - z_2)S_F(z_1 - x_3)S_F(x_4- z_1)]
\end{eqnarray}

Bath members in (92) are equal after the substitution $z_1 \leftrightarrow z_2$. Therefore,

\begin{eqnarray}
\tau (x_1, x_2,x_3,x_4) =ie^2 \int dV_{z_1}dV_{z_2}\tilde {\gamma}(z_2)\tilde {\gamma}(z_1) D_F(z_2 - z_1) S_F( z_1 - x_2 )S_F( x_1 -z_1 ) \times \nonumber \\
\times  S_F(z_2 - x_3)S_F(x_4- z_2) 
\end{eqnarray}

Thus, we state the result corresponding to  classical quantum electrodynamics. However, instead of usual Green functions of QED it should be applied  Green functions for curved spacetime .

To find the scattering matrix, we substitute the relation (93) in the expression (89) taking into account that $x_1=x, x_2=z, x_3=x', x_4=z'$:

\begin{eqnarray}
    <p's';q'r'  \, out|ps;qr,\,in>  =   \delta_{ij} - ie^2 \int dV_{z_1}dV_{z_2}\tilde {\gamma}(z_2)\tilde {\gamma}(z_1)\int  dV_x \int  dV_{x'} \int  dV_z \int dV_{z'} \times \nonumber\\
  \times    {\bar U}_{p's'}(x')(i \tilde{\gamma}(x') \cdot\tilde {\nabla}_{x'}   -m_{\mu}){\bar V}_{qr}(z)(i\tilde{\gamma}(z) \cdot \tilde {\nabla}_{z}   -m_e)\times \nonumber\\
 \times D_F(z_2 - z_1) S_F(z_1 - z)S_F(x-z_1)S_F(z_2-x')S_F(z'- z_2) \times \nonumber\\
 \times [\stackrel{\longleftarrow}{(-i\tilde{\gamma}(z') \cdot  \tilde {\nabla}_{z'}  -m_{\mu})} V_{q'r'}(z')] \stackrel{\longleftarrow}{(-i \tilde{\gamma}(x) \cdot  \tilde {\nabla}_{x}  -m_e)} U_{ps}(x)
\end{eqnarray}

We have the relations:

\begin{equation}
(i \tilde{\gamma}(z)\cdot \tilde {\nabla}_{z}   -m_{e}) S_F(z_1 - z)=\delta (z_1 - z),
\end{equation}

\begin{equation}
(i\tilde{\gamma}(x') \cdot \tilde {\nabla}_{x'}   -m_{\mu}) S_F(z_2-x')=\delta (z_2 - x'),
\end{equation}

\begin{equation}
S_F(z'-z_2) \stackrel{\longleftarrow}{(-i\tilde{\gamma}}(z') \cdot \tilde {\nabla}_{z'}   -m_{\mu})= \delta (z'-z_2),
\end{equation}

\begin{equation}
S_F(x - z_1) \stackrel{\longleftarrow}{(-i\tilde{\gamma}}(x)\cdot\tilde {\nabla}_{x}   -m_e)= \delta ( x - z_1 ).
\end{equation}

Taking into account (95) - (98), we find that

\begin{eqnarray}
    S_{fi}  =  - i{e^2} \int  dV_{z_1} dV_{z_2} \bar  {V}_{qr}(z_1) \tilde {\gamma}_{\mu} (z_1) U_{ps}(z_1) D^{\mu\nu}_F(z_2 -z_1) \bar {U}_{p's'}(z_2)\tilde {\gamma}_{\nu}(z_2) {V}_{q'r'}(z_2). 
\end{eqnarray}

For  Minkowski space we can present  $U_{ps}, \bar {U}_{p's'},\bar {V}_{qr}, {V}_{q'r'}$ in terms of plane waves.
Substituting in (99) 

\begin{equation}
D^{\mu\nu}_F(z_1 -z_2) = \frac {1}{{(2\pi)}^4}\int dq \, exp (-iq(z_1 -z_2)) \tilde D^{\mu\nu}_F(q),
\end{equation}
where $\tilde D^{\mu\nu}_F(q)$ is the Fourier transform for $D^{\mu\nu}_F(z)$, we shall find that all signs of integrals in (99) vanish, and we shall obtain
the usual expression  for the annihilation of electron-positron pair   (see \cite {Pes} sec.5.1). 

\subsection{The generalized Feynman rules}

On the base of results of sec. 5.2 and 5.3 we can formulate the generalized Feynman rules for  quantum electrodynamics in curved spacetime
\footnote{We recall that, in the general case, the solutions $U_{ps}(x), V_{ps}(x)$ for out-states will not be equal to the corresponding solutions for in-states.}:

\begin{enumerate}

\item  For each internal fermion line  the factor $iS_F(x-y)$.

\item   For each internal photon line the factor $iD_F(x-y)$.

\item  For each vertex the factor

\begin{equation}
 -ie \int dV_z \tilde {\gamma}(z).
\end{equation}

\item   For each external photon line with  momentum $k$ and polarization $\lambda$ the factor $A_{k\lambda} (x)$
in  the case of in-line and the factor  $A_{k\lambda}^{\ast} (x)$ for out-line.

\item   For each external fermion line with  momentum $p$ and spin $s$ the factor $U_{ps} (x)$ in  the case of in-line and the factor $\bar {U}_{ps} (x)$
for out-line.

\item   For each external antifermion line with  momentum $p$ and spin $s$ the factor  $\bar {V}_{ps} (x)$ in  the case of in-line and the factor 
 $V_{ps} (x)$ for out-line.

\end{enumerate}

For Minkowski space these rules are  the Feynman rules in \cite {Pes}, Appendix A1.

\section{Conclusion}
\begin{enumerate}
\item We have found the generating functional for the interaction of Dirac and electromagnetic fields.
\item  Using reduction formalism, we have calculated the scattering matrices for elementary processes of quantum electrodynamics in curved spacetime:
Compton scattering and the annihilation of electron-positron pair (for the tree-level approximation). 
\item On the base of these results we have formulated the generalized Feynman rules for  the electromagnetic interaction in curved spacetime.
\item Another electrodynamic processes can be studied by means of these Feynman rules and crossing-symmetry.
\end{enumerate}

\end{document}